\documentclass[a4paper,11pt]{article}
\usepackage{epsfig}
\usepackage{rotating, graphicx, color,tabularx}

\newcommand{\ie}{{\it i.e.\ }}
\newcommand{\etal}{{\it et al.\ }}
\renewcommand{\baselinestretch}{1.5}
\begin{document}

\title{eBay Users form Stable Groups of Common Interest}
\author{J\"org Reichardt and Stefan Bornholdt\\
Institute for Theoretical Physics, University of Bremen, 
28359 Bremen, Germany\\ \texttt{\{reichardt,bornholdt\}@itp.uni-bremen.de}}
\maketitle
\clearpage
\begin{abstract}
\noindent Market segmentation of an online auction site is studied by analyzing the users' bidding behavior. The distribution of user activity is investigated and a network of bidders connected by common interest in individual articles is constructed. The network's cluster structure corresponds to the main user groups according to common interest, exhibiting hierarchy and overlap. Key feature of the analysis is its independence of any similarity measure between the articles offered on eBay, as such a measure would only introduce bias in the analysis. Results are compared to null models based on random networks and clusters are validated and interpreted using the taxonomic classifications of eBay categories. We find clear-cut and coherent interest profiles for the bidders in each cluster. The interest profiles of bidder groups are compared to the classification of articles actually bought by these users during the time span 6-9 months \textit{after} the initial grouping. The interest profiles discovered remain stable, indicating typical interest profiles in society. Our results show how network theory can be applied successfully to problems of market segmentation and sociological milieu studies with sparse, high dimensional data.
\end{abstract}
\clearpage
\section*{Introduction}
\noindent The internet has changed the way people communicate, work, and do business. One example are online auction sites, the largest being eBay with its more than $150$ million registered users world wide \cite{Economist}. An interesting aspect of eBay's success is its transparency. The market is fully transparent as the trading history of every user is disclosed to everyone on the internet. We here study the relationship between the participants of this market. 

\begin{figure}[h]
\begin{center}
	\includegraphics[width=5cm]{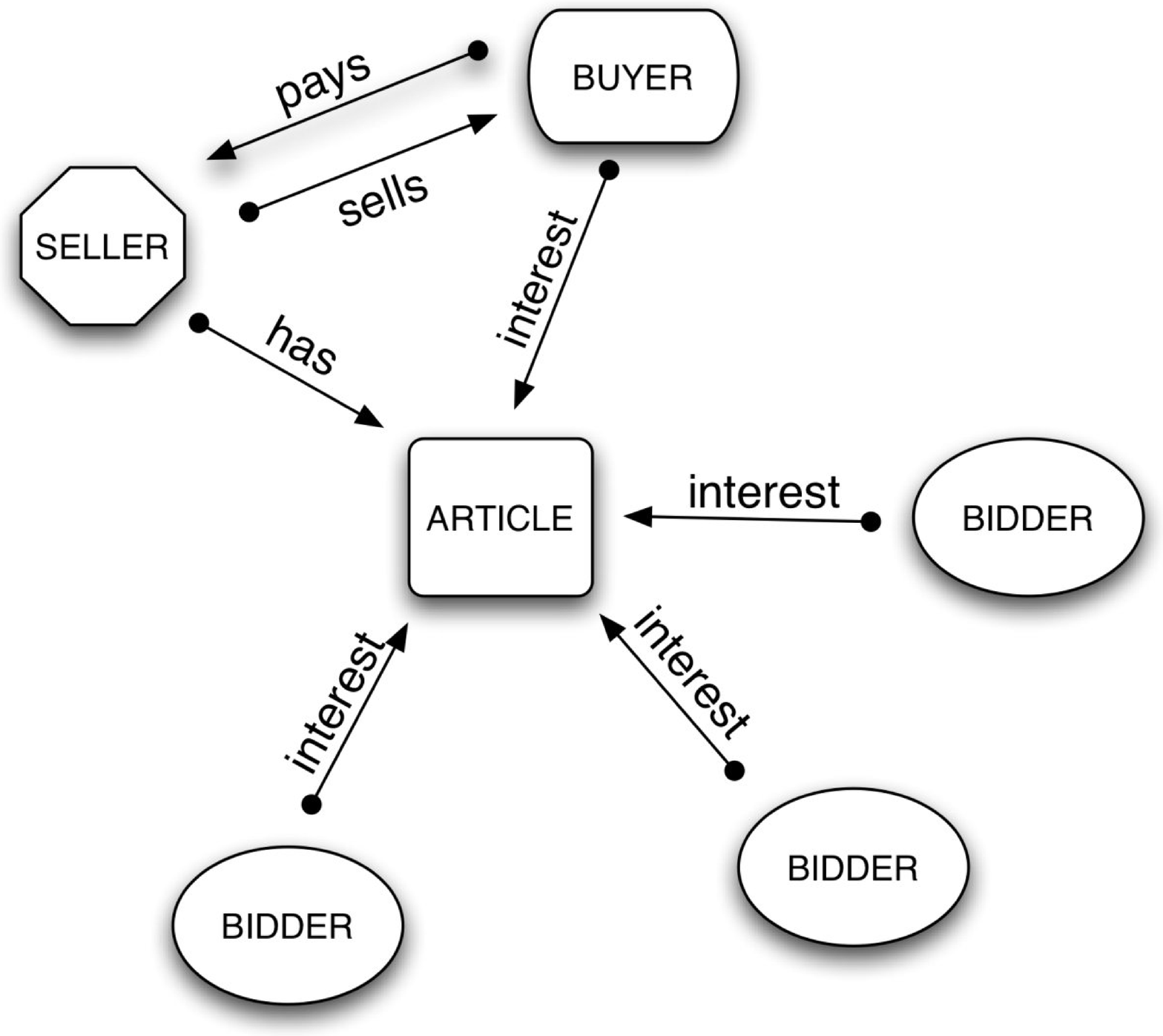}
\end{center}	
	\caption{Structure of a single auction. Users express their common interest in a particular article by bidding. The user with the highest bid wins the auction and exchanges money and the article with the seller. EBay earns a fee with every transaction. Users of the auction site, \textit{i.e.} bidders, buyers or sellers, may change their role in a different auction of another article.}
	\label{EBayPrinciple}
\end{figure}

Let us first recall the operating principle of an online auction in Figure \ref{EBayPrinciple}. Users may offer goods through the online platform and set a deadline when their auction will end. Articles are listed under a certain taxonomic product category by the seller and are searchable platform wide. Users with a particular demand either browse through the articles listed in an appropriate category or search for articles directly. Until the end of the auction they may bid on the article. The user with the highest bid at the end of the auction wins (so called hard-close) and buys the article. In every new auction, users may assume different new roles as sellers, bidders or buyers. The market can be represented as a graph with the users and/or articles as the nodes and the links denoting their interactions as shown in Figure \ref{EBayPrinciple}.   

A number of researches have presented statistical studies of trading \cite{Kahng1} and analyses of bidding strategies and ending auction ending rules \cite{Ockenfels1, Kahng3}. In this contribution we focus on the market segmentation of the eBay auction site. Our approach is based on the assumption that at a certain level of abstraction the population of consumers can be separated into relatively clear-cut and homogenous sub-groups corresponding to certain customer milieus or market segments \cite{WassermanFaust}. We assume that customers of the same type are described by a common pattern in their consumer interests which leads to a higher probability of bidding for the same article \cite{NewmanMixing}.   

In particular, we perform a cluster analysis \cite{JainReview, ArabieHubert} of the bidding behavior of about one million users. Groups of eBay users with common interest or demand are detected using solely the information of which users competed in the same auctions. The classification is based on a very sparse and high dimensional data set \cite{ClusterChallenges} with only slightly more than $3$ auctions per bidder on average (out of $1.6$ million possible auctions). Conventional analysis techniques such as correspondence analysis \cite{WassermanFaust, CorrespondenceAnalysis} have to make use of a similarity measure between articles in order to reduce the dimensionality and coarse-grain the data, such as exploiting the annotation of articles into product categories. However, this bears several pitfalls: First, the annotations are defined by the seller who lists the article such that it can be found efficiently, hence, the categorization is mainly a taxonomy. Using this to coarse-grain the data would introduce a bias in the analysis. Second, eBay categories differ largely in size when counting the number of articles in the category as well as the number of sub-categories. Correcting for this again may introduce a bias. Third, using the category taxonomy for coarse-graining induces a hierarchy in the data, as all articles below the cut in the taxonomy tree are subsumed. Fourth and most importantly, it is not clear at which level in the category tree a coarse-graining should be performed and whether this level should be the same for all branches. 
 
Our analysis is independent of taxonomic categories and dimensionality reduction. It allows for hierarchical and overlapping cluster structures, and we find evidence for both. The product categories are solely used to interpret the results of our study, \ie provide interest profiles of user groups found in terms of this taxonomy.    

By clustering users directly according to a common demand spectrum, we also circumvent problems of conventional basket analysis done by frequent item sets \cite{Agrawal1994, Hipp2000,HyperCluster,Sarwar1}. The latter asks which articles are frequently demanded by a single person. This analysis is performed for all articles averaging over the entire population of consumers and hence results in the least common denominator of articles which may then be bundled together and marketed together to the whole population of customers.  The same is true for cluster analysis of eBay categories \cite{Kahng3}. The proposed \textit{network cluster analysis}, however, reveals information about people and their diverse and possibly very special interests. 

\section*{Dataset}
\noindent A dataset consisting of over $1.59$ million auctions was obtained from the German eBay site \texttt{www.ebay.de} ending during the pre-Christmas season December 6$^{th}$ and 20$^{th}$ 2004. Considering only articles with locations in Germany, we recorded the user-id of seller, buyer, and all bidders competing in each auction, as well as the individual bids and the product category in which the article was listed (excluding articles listed in the real estate category which was in a beta testing phase at the time). Since auctions last between $7$ and $10$ days depending on the choice of the seller, we thus cover a bidding period of up to 25 days. We believe the pre-Christmas time is a suitable time for analysis for the following reasons: First, traffic is very high. In fact, there was a broad advertising campaign in Germany advertising to shop for Christmas presents on eBay. Second, we only considered auctions and expect that users are unlikely to bid for articles for which they cannot assess a fair price. Third, if users shop for presents, then we can gain some information about their family background, e.g. people shopping for toys will most likely have a child themselves or among their closer relatives. Our findings indicate that this is indeed the case. Table \ref{DataSet} summarizes the dataset in its basic parameters. There are far less sellers than bidders and only $38\%$ of the sellers  also act as bidders or buyers. This indicates that users are split into those mostly selling and those mostly buying.  

\begin{table}[b]
\caption{Summary of the data set of online auctions obtained between Dec. 6$^{th}$ and 20$^{th}$ 2004. Numbers in millions. }
\label{DataSet}
	\begin{center}
	\begin{tabular}{rl}
	\hline
	auctions observed: & $1.59$\\
	users acting as buyer: & $0.95$\\
	users acting as seller: & $0.37$\\
	users acting as bidder: & $1.91$\\
	users acting as seller and bidder: & $0.14$\\
	users acting as seller and buyers: & $0.08$\\
	\hline
	\end{tabular}
	\end{center}
\end{table}

\section*{User Activity and User Networks}
The activity of the users is measured via the probability mass distributions of the number of articles sold $p(s)$, bought (auctions won) $p(w)$, and bid on $p(a)$. Though it is possible to bid multiply in a single auction, we neglect this fact and use ``bid'' and ``take part in an auction'' synonymously. Similar to previous studies \cite{Kahng1}, we find fat tailed distributions of the user activity in the form $p(x)\propto x^{-\kappa}$. For the number of bidders $b$ taking part in an auction, the ``attractiveness of an article'', we find an exponential distribution $q(b)\propto \alpha^b$. Table \ref{DataSetActivities} summarizes the parameters obtained by maximum likelihood fitting for these distributions \cite{PowerlawYen, NewmanPowerlaw}. Plots of the data can be found in the supporting online material.   

The fat tails of the distribution are striking given the short time span observed. Consider the most active bidder taking part in over $800$ auctions! This user seems to follow a gambling strategy bidding only minimal amounts as he/she wins only a few of these auctions. The most successful buyer who won 201 auctions on the other hand took part in only 208 auctions. This hints at a diversity of strategies employed by users of the online auction site. Curiously, the article most desired and attracting $39$ different bidders was a ride in a red Coca-Cola-Truck. 
   
\begin{table}
\caption{Activity distributions of observed users in auctions. Shown are the average values and the exponents $\kappa$ of the tail of the distribution ($x\geq 10$) if they follow a power law $p(x)\propto x^{-\kappa}$ or the parameter $\alpha$ if the distribution has an exponential form $q(x)\propto \alpha^x$. All parameter estimates are maximum likelihood estimates. }
\label{DataSetActivities}
	\begin{center}
	\begin{tabular}{rlll}
	\hline
	& $\langle x\rangle$ & $\kappa$ & $\alpha$\\
	\hline
	articles sold per seller: & $4.3$ & $2.37$ & \\
	auctions taken part in per bidder: & $2.9$ & $2.78$ &\\
	articles bought per buyer: & $1.7$ & $3.38$ & \\
	bidders per auction: & $3.4$ & & $0.71$\\
	\hline
	\end{tabular}
	\end{center}
\end{table}

From the original data a number of market networks can be constructed, such as the network of users connected by actual transactions, or the network of sellers that are connected if they have sold to (or received bids from) the same user. Then, the links in the network would represent competition or a possibility for cooperation, depending on the portfolio of articles offered by these sellers. 

Here, we focus only on the bidder network based on single articles. Two bidders are linked if they have competed in an auction. Since all users that bid in a single auction are connected, this network results from overlaying fully connected cliques of bidders that result from each auction. Such graphs are also known as affiliation networks \cite{StrogatzNature, NewmanHighCluster,NewmanRG}. 

Prior to a cluster analysis in this bidder network, we study its general statistical properties looking for indications of cluster structure \cite{NewmanRGPNAS}. We compare the results to a randomized null model (RNM) obtained from reshuffling the original data, \ie keeping the attractiveness of each auction and the activity of each bidder constant, but randomizing which bidders take part in which auction. 

\begin{figure}[t]
\includegraphics[width=5.5 cm]{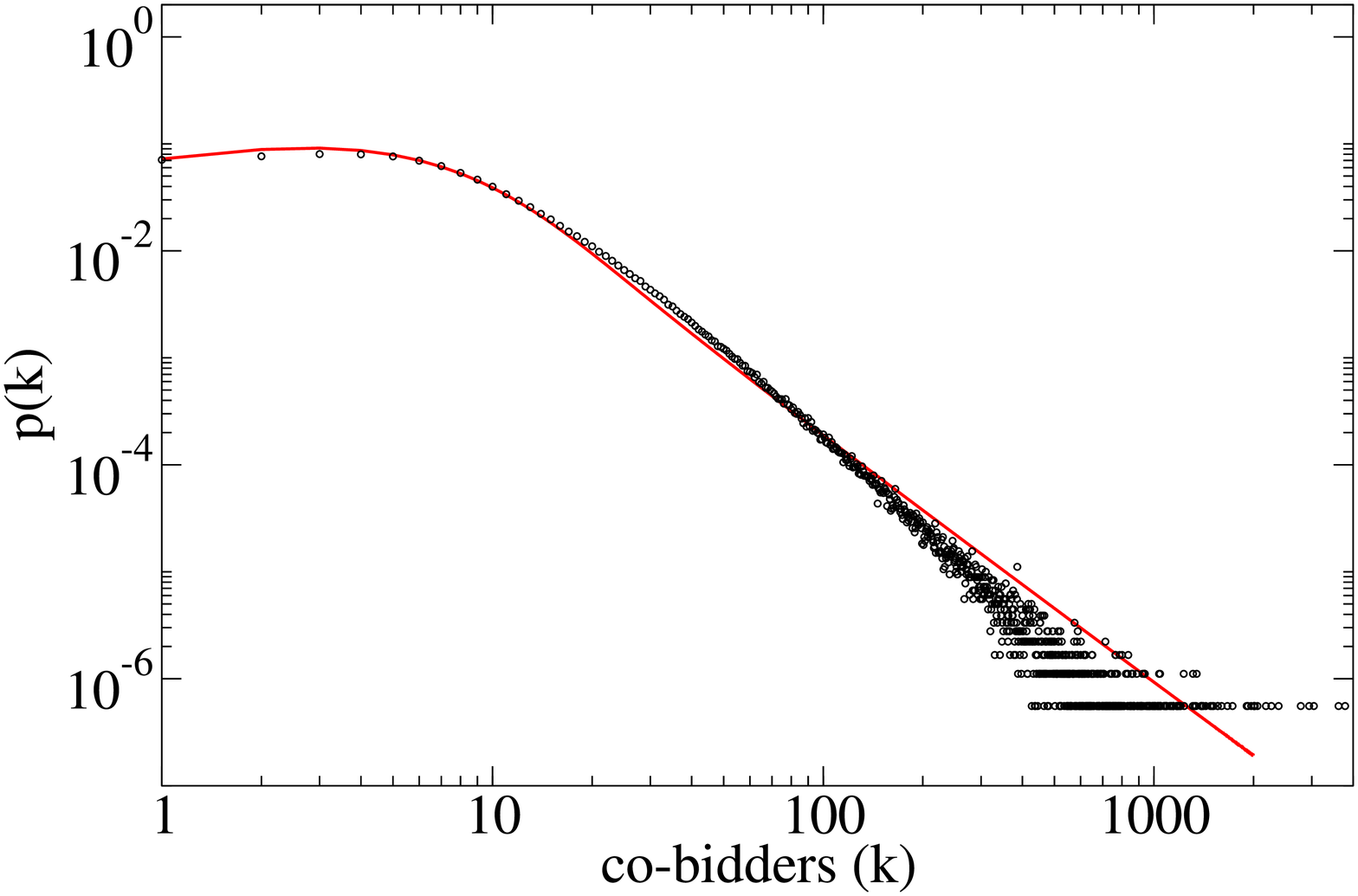}
\includegraphics[width=5.5 cm]{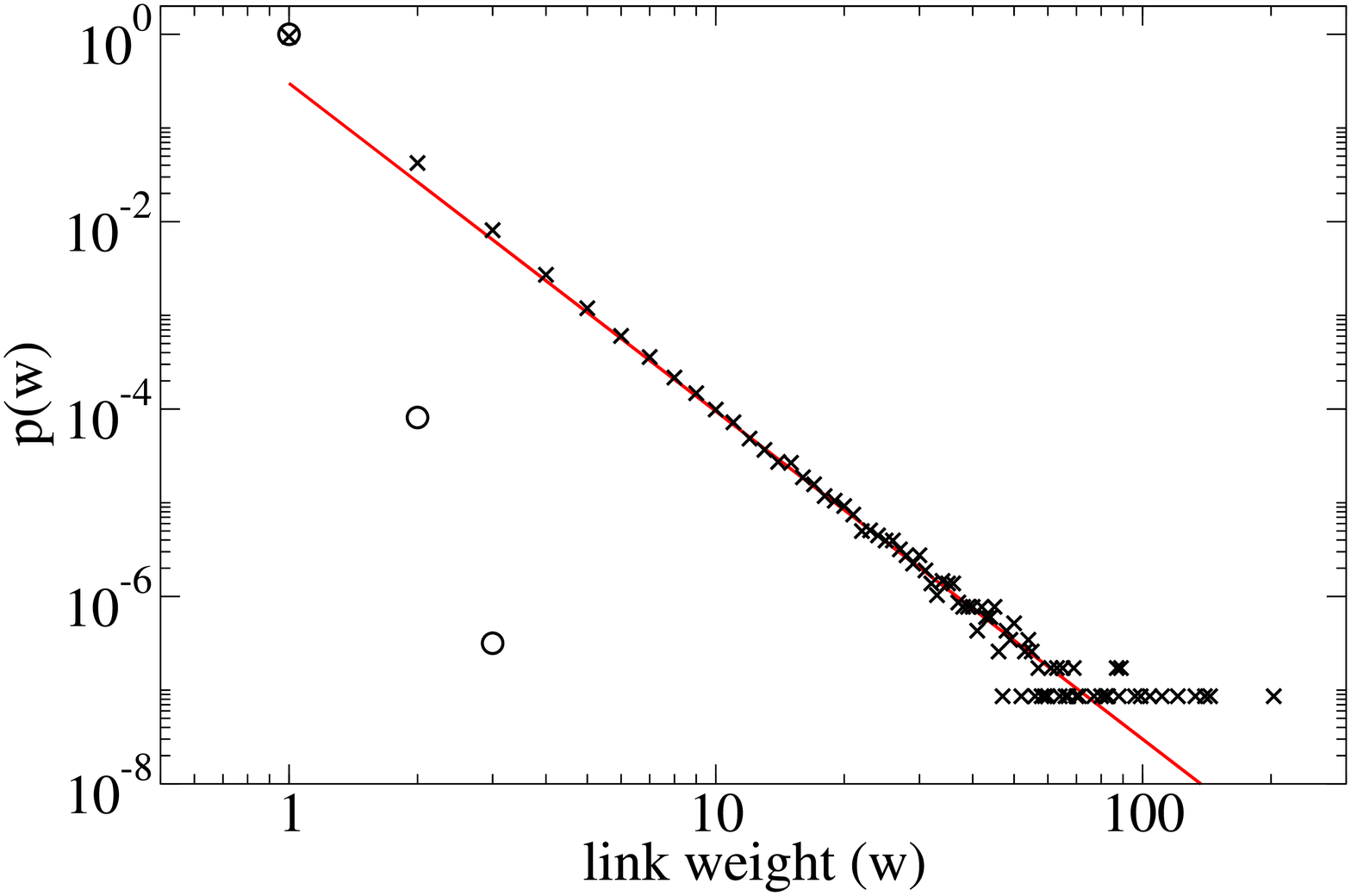}
\includegraphics[width=5.5 cm]{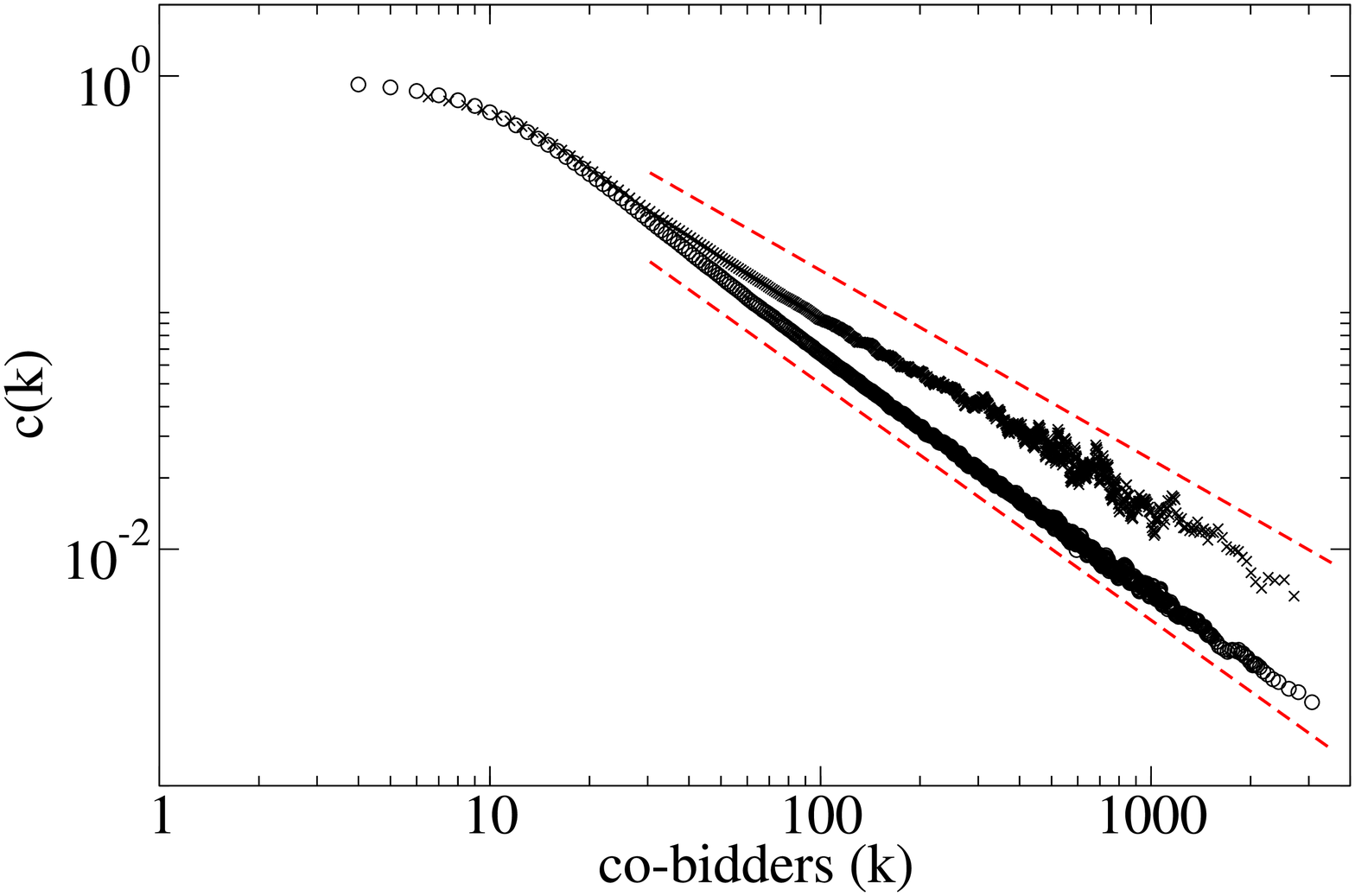}
\caption{Comparison of the bidder network with a random null model (RNM). Left: degree distribution. The solid line represents a theoretical degree distribution according to equation (\ref{DegreeDist}). Middle: distribution of the link weights in the bidder network. (o) for the RNM and (x) for the empirical data. Right: Distribution of the clustering coefficient $c(k)$ as a function of the degree $k$ of the nodes. (o) for the RNM and (x) for the empirical data. The two dashed lines indicate two power laws $\propto k^{-\kappa}$ with exponents $\kappa=1$ and $\kappa=0.8$, respectively.}
\label{RNMComparison}
\end{figure}

Furthermore, we compare the bidder network with theoretical predictions we can derive from the distribution of the bidding activity $p(a)$ and the distribution of the attractiveness of auctions $q(b)$. The degree distribution of the bidder network can be calculated from $p(a)$ and $q(b)$ using a generating function formalism  \cite{GeneratingFunctions,NewmanRG} assuming that bidders never meet twice in different auctions. With a power-law form of $p(a)\propto a^{-\kappa}$ and an exponential form for $q(b)\propto\alpha^b$ as before, the degree distribution in the bidder network amounts to:
\begin{equation}
p(k) = \alpha^k\sum_ap(a)(1-\alpha)^{2a}\left(\begin{array}{c}2a-1+k\\ k\end{array}\right).
\label{DegreeDist}
\end{equation}
Figure \ref{RNMComparison} shows a comparison of the empirical data from the bidder network and the theoretical curve (\ref{DegreeDist}). The shapes of the distributions agree quite well, given that we use an estimate based on only two parameters $\kappa=2.78$ and $\alpha=0.71$. 

A theoretical expectation for the average number of neighbors in the bidder network is be given by $\langle k\rangle=2(\langle b\rangle-1)\langle a\rangle=14$ where $\langle b\rangle$ is the average number of bidders per auction and $\langle a\rangle$ is the average number of auctions taken part in by a bidder. This estimate is in excellent agreement with the result from the RNM, but larger than in the actual data, indicating that the probability to meet in an auction twice is not zero confirming our expectation. See Table \ref{BidderNetwork} for a summary of the basic parameters of the empirical data and the RNM.

Comparing the distribution of the link weights, \ie the number of times two bidders have met in an auction, we find a much more prominent difference between the data and the RNM. Figure \ref{RNMComparison} shows that the weights of the links in the bidder network are distributed with a power law tail. Approximately $6\%$ of all links correspond to pairs of bidders which have met more than once. If there would be no common interest among bidders, practically all links would have weight 1 as is indeed the case for the RNM.  

Additionally to the degree distribution, we compare the distribution of the clustering coefficient as a function of the degree of a node. The clustering coefficient $c(k)$ denotes the average link density among the neighbors of a node of degree $k$. Due to the construction process of the network as an affiliation network, we expect that for large numbers of neighbors $k$ the clustering coefficient $c(k)$ scales as $k^{-1}$ in case of random assignment of bidders to auctions \cite{NewmanHighCluster}.  Figure \ref{RNMComparison} shows that this is indeed the case for the RNM, but the actual data deviates strongly for bidders with a large number of neighbors. This effect can arise from two processes: either bidders with whom one competes in two different auctions also meet independently in a third auction, or that there is an increased probability that one will compete again with a bidder one has already met once in an auction. Both explanations support our assumption of the presence of clusters of users with common interest. 

With these comparisons, we have shown that the bidder network is far from randomly constructed and we will proceed by studying the cluster structure for which we have found indirect evidence already.  

\begin{table}[t]
\caption{Summary of basic parameters for the bidder network with two bidders linked, if they have competed in an auction. Shown are the actual data, the parameters for a random null model (RNM) and the reduced version of the network used for cluster analysis.}
\label{BidderNetwork}
	\begin{center}
	\begin{tabular}{rlll}
	\hline
	& data & RNM & reduced\\
	\hline
	number of nodes: & $1.8\times10^6$ &$1.8\times10^6$ & $0.9\times10^6$\\
	number of links: & $11.6\times10^6$ & $12.6\times10^6$& $7.4\times10^6$\\
	average degree: & $12.9$ & $13.9$ & $16.4$\\
	assortativity: & $0.02$ & 0.0 & 0.03\\
	\hline
	\end{tabular}
	\end{center}
\end{table}

\section*{Market Segmentation}
\subsection*{Network Clustering}
\noindent The analysis of the user interests in the eBay market is based on the bidder network as constructed in the previous section. The links in this network represent articles the connected bidders (nodes) have a common interest in. We reduce the network to only those bidders that have taken part in at least two auctions and we consider only auctions with a final price below $1,000$ Euro, thereby focussing on consumer goods. See Table \ref{BidderNetwork} for the basic parameters of this reduced network. 

If we now find groups of users (clusters or communities \cite{Girvan,NewmanQPNAS,GuileraReview}) with a high density of links among themselves and a low density of links to the rest of the network, the total set of links within such a group of users can be interpreted as a unifying common interest of this group. We assign the users into communities as to maximize a well established quality function known as network modularity $Q$ defined by Girvan and Newman (GN) \cite{Girvan03}. The definition of $Q$ can also be written as \cite{ReichardtPRE}:
\begin{equation}
MQ=\sum_s\underbrace{\left(m_{ss}-\gamma[m_{ss}]\right)}_{c_{ss}}=-\sum_{s<r}\underbrace{\left(m_{rs}-\gamma[m_{rs}]\right)}_{a_{rs}}.
\label{Modularity2}
\end{equation}
Here, the first sum runs over all group indices $s$, while the second over all pairs of different group indices $s>r$, $m_{ss}$ is the number of internal links in group $s$ and $[m_{ss}]$ is an expectation value for this quantity in case of a random assignment of bidders into groups and is given by $[m_{ss}]=K_s^2/4M$. By $K_s$ we denote the total number of links emanating from members of group $s$ and $M$ is the total number of links in the network. Equivalently, $m_{rs}$ is the number of links between members of group $r$ and $s$ and $[m_{rs}]$ is the corresponding expectation values given by $[m_{rs}]=K_sK_r/2M$ \cite{ReichardtPRE}.  $Q$ is maximal, when the sum of cohesions $c_{ss}$, defined as the difference between the actual and expected number of within group links, is maximal. Equivalently, $Q$ is maximal when there are many less links between different groups than expected for a random assignment of nodes into communities, \ie the sum of adhesions $a_{rs}$ is minimal. Note that any assignment of bidders into groups which maximizes $Q$ will be characterized by both, maximum cohesion of groups, and minimal adhesion between groups. If $Q$ is maximal, every node is classified in that group to which it has the largest adhesion, otherwise it could be moved to a different group to increase $Q$. Additional to the original definition of $Q$ by GN, we have introduced a parameter $\gamma$ which allows to adjust the relative influence of actually present and expected links in the definition (setting $\gamma=1$ recovers the original definition of GN). Values of $\gamma$ smaller or greater than one lead to larger or smaller communities, respectively. Comparing classifications obtained at different values of $\gamma$ allows the detection of hierarchy and overlap in the cluster structure. See Ref. \cite{ReichardtPRL,ReichardtPRE} for examples and further details of this variation.      

The technical details of how the bidders can be assigned into groups such that $Q$ is maximized are given in Refs. \cite{ReichardtPRE,NewmanQPNAS}. We allowed for maximally $500$ different groups of bidders in our analysis which gives a sufficient level of detail.

\begin{figure}
\begin{tabular}{cc}
	\begin{tabular}{l}
	\fbox{\includegraphics[width=7.5cm]{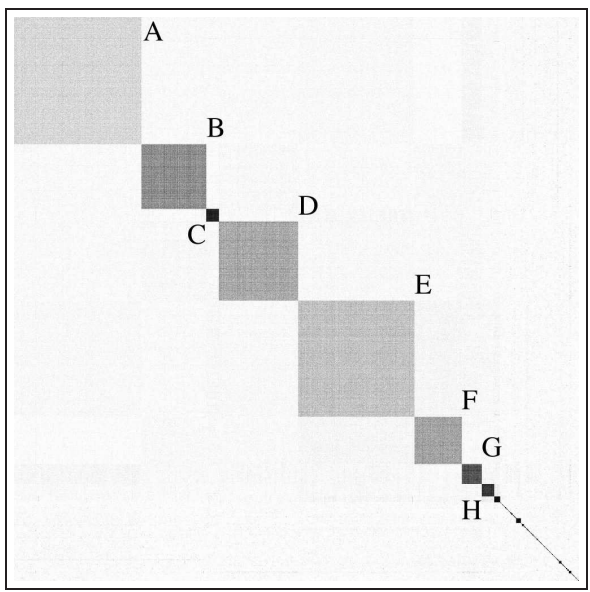}}\\
	\fbox{\includegraphics[width=7.5cm]{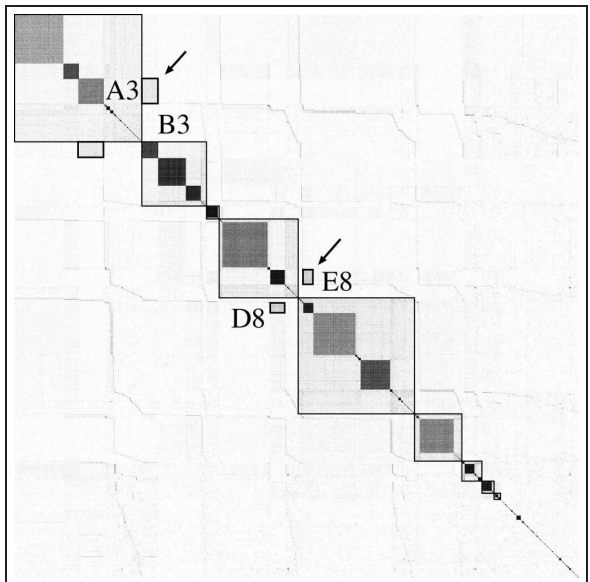}}\\
	\fbox{\includegraphics[width=7.5cm]{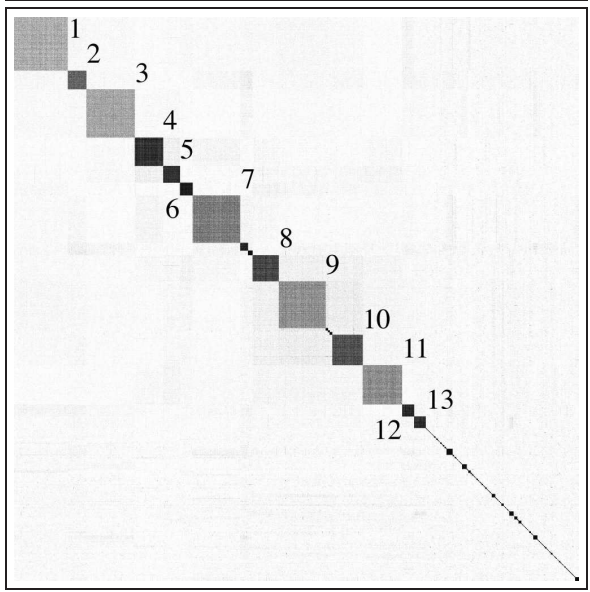}}
	\end{tabular} &
	\begin{tabular}{l}
	\fbox{\includegraphics[height=8.1cm, angle=90]{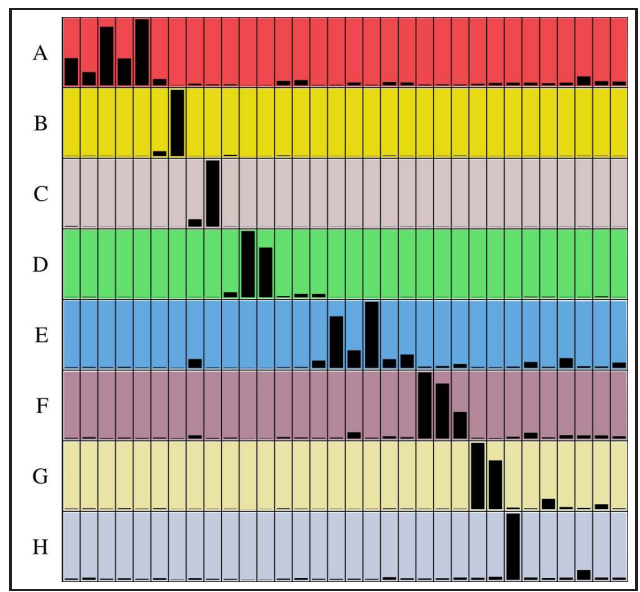}}\\
	\fbox{\includegraphics[height=8.1cm, angle=90]{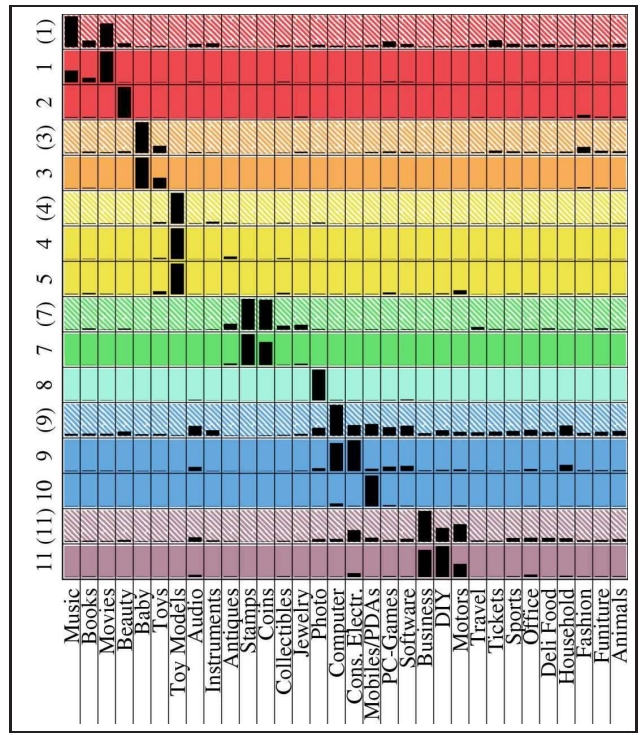}}\\
	\begin{minipage}[t][5.5cm][t]{8cm}
	 \renewcommand{\baselinestretch}{1}
	\caption{\small\textbf{Left:} N$\times$N adjacency matrix of the bidder network in three different orderings. A pixel in row $i$, column $j$ corresponds to an auction in which bidder $i$ and $j$ have competed. Shown are $N=889828$ bidders (nodes) and $M=7373008$ pairwise competitions (links). Grey levels correspond directly to link density in this network and hence to the probability of competing in an auction. Top: $\gamma=0.5$ ordering, bottom: $\gamma=1$ ordering and middle: consensus ordering of top and bottom. \textbf{Right:} Odds ratios of bidding in one of the 32 main eBay product categories for classified users. Top: from $\gamma=0.5$ classification, bottom: from $\gamma=1$ classification. Spectra with a dashed background (cluster id in parenthesis) show customer purchases $6-9$ months after original classification. See text for details.}
	\label{Clusterings}
	\end{minipage}
	\end{tabular}
\end{tabular}
\end{figure}

The left part of Figure \ref{Clusterings} compares the results obtained with $\gamma=0.5$ and $\gamma=1$. Shown are the adjacency matrices $A_{ij}$ of the largest connected component of the bidder network. A black pixel at position $(i,j)$ and $(j,i)$ is shown on an $889828\times889828$ square if bidders $i$ and $j$ have competed in an auction and hence $A_{ij}=1$, otherwise the pixel is left white corresponding to $A_{ij}=0$. The rows and columns are ordered such that bidders who are classified as being in the same group are next to each other. The internal order of bidders within groups is random. The order of the groups was chosen to optimally show the correspondence between the ordering resulting from the $\gamma=0.5$ and the $\gamma=1$ ordering. In this representation, link densities correspond to pixel densities and thus to grey levels in the figure. Information about the exact size and link density contrast of the clusters is given in the supporting online material. Note the high contrast between internal and external link density. 

At the top of Figure \ref{Clusterings}, we show the adjacency matrix ordered according to an optimal assignment of bidders into groups with $\gamma=0.5$. Clearly, a small number of major clusters of bidders and a large number of smaller clusters are identified, strongly connected internally and well separated from one another. The largest $8$ clusters are marked with letters A through H. Of all bidders in the network,  $85\%$  are classified in these 8 clusters. At the bottom, we show the same adjacency matrix, but now rows and columns are ordered according to an optimal assignment of bidders into groups with $\gamma=1$. As expected, we find a larger number of smaller, denser clusters which are numbered 1 through 13. In order to analyse whether the network has a hierarchical or overlapping cluster structure, we define a consensus ordering of the bidders from the $\gamma=0.5$ and $\gamma=1$ ordering by reshuffling the internal order of the $\gamma=0.5$ clusters according to the $\gamma=1$ clustering. Remember the orderings for the two values of $\gamma$ were obtained independently. If the network possesses a hierarchical structure in the sense that the clusters obtained at higher values of $\gamma$ lie completely within those obtained at lower values of $\gamma$, then the consensus ordering would not differ from the ordering at $\gamma=1$. If, however, clusters at lower values of $\gamma$ overlap and this overlap forms its proper cluster at higher values of $\gamma$, the network is not entirely hierarchical. These aspects will become immediately clear by looking at the middle part of Figure \ref{Clusterings}. For clarity, we have marked the borders of the $\gamma=0.5$ clustering. Clusters 1 and 2 fall entirely within cluster A giving an example of a cluster hierarchy. Cluster 3, however, is split by the consensus ordering into one part A3 belonging to A, and B3 belonging to B (see arrows in figure). It is now clear that clusters A and B actually have some overlap which was not visible in the $\gamma=0.5$ ordering. This overlap is concentrated in cluster 3, parts of which belong stronger to either A or B. Clusters 4 and 5 then fall again completely within cluster B. Clusters 6 and C are practically identical. Cluster D has a number of sub-clusters, the largest of which is 7 and overlaps with cluster E through cluster 8 as before (see arrows again). Group E has two more sub-groups 9 and 10 while clusters 11, 12 and 13 fall entirely within clusters F, G and H, respectively. More details about hierarchical and overlapping cluster structures including some toy examples can be found in \cite{ReichardtPRE}. 

\subsection*{Cluster Validation, Interpretation and Time Development}
\noindent To validate the statistical significance and to rule out the possibility the observed cluster structure is merely a product of the clustering algorithm or the particular method of constructing the network from overlapping cliques of bidders, we compare the results to those obtained for appropriate random null models.  Maximizing $Q$ also for the RNM version of the bidder network, again taking into account those bidders which took part in at least two auctions, we find a value of $Q=0.28$ at $\gamma=1$ which is significantly less than the value of $Q=0.64$ for the empirical data. Furthermore, the RNM shows all equal sized clusters, while the real network clearly possesses major and minor clusters. A random graph with the same number of nodes and links, \ie disregarding the scale free degree distribution and the affiliation network structure of the graph, would yield only $Q=0.23$ \cite{ReichardtPRE}.
 
Until now we have only found groups of bidders that have am increased probability to meet other members of their groups in the auctions they take part in. The eBay product categories are now used in order to find an \textit{interpretation} for the common interests that lead to the emergence of the cluster structure of the bidder network. Since cluster sizes vary and the number of articles in the individual categories is very diverse, we calculate the odds ratios (OR) for bidding in one of the 32 main categories. This odds ratio is defined as
\begin{equation}
OR_{Cs}=\frac{P(\mbox{bidding in $C$} | \mbox{member of cluster $s$})}{P(\mbox{bidding in $C$} | \mbox{not member of cluster $s$})},
\end{equation}
\ie the ratio of the odds of  bidding in category $C$, given a bidder is member of group $s$ vs.\ the odds of bidding in category $C$ given the bidder is member of any group $r\neq s$. The right hand side of Figure \ref{Clusterings} shows a graphical representation of the odds ratios for clusters $A$ through $H$ and most of the clusters $1$ through $13$. All spectra are normalized. The exact numerical values can be found in the supporting online material. Clusters from the $\gamma=1$ assignment are more specific with less entries in the category spectrum and larger ORs. 

Cluster A unites bidders interested in articles listed in the baby, beauty, fashion, books, movies and music category. Cluster 1 then represents a  more specifically content oriented user group mainly interested in books, movies and music. As we have seen, cluster 1 is an almost complete sub-cluster of A. Cluster 2 is also a complete sub-cluster of A and encompasses bidders mainly interested in cosmetics and fashion. 

Cluster B contains two sub-clusters 4 and 5, both annotated in the toy model category. Closer inspection, however, reveals that cluster 4 is mainly characterized by its interest in model railways while the bidders in cluster 5 have a passion for model cars, radio controlled models, slot cars and the like. Note the advantage of clustering based on single articles. The clusters we find with one simple unbiased method combine top level categories as in the case of cluster 1 or can only be described be resorting to sub-categories as in the case of clusters 4 and 5. From the left part of Figure \ref{Clusterings}, we had observed that cluster 3 is responsible for a large part of the overlap between clusters A and B. We see that users in this group 3 have their main interests in the baby and toy category. The overlap of cluster A and B is hence mediated via the toy category. Members of cluster A and B mainly meet in toy auctions. The interpretation of the other clusters is then equally straightforward. 

Bidders in clusters C and the practically identical cluster 6 take interest in audio equipment and instruments. Cluster D represents bidders with an inclination to collecting, their bids being placed in the antiques, jewelry, stamps and coins category (cluster 7). The bidders in cluster E are mainly shopping for technological gadgets, computers, consumer electronics, software, mobile phones, PDAs etc.\ (clusters 9 and 10). Their overlapping interest with bidders from cluster D is in items from the photo category (cluster 8). In groups F and 11, we find predominantly practically oriented users who place their bids mainly in the categories of automotive spare parts, business and industry (where a lot of tools and machinery are auctioned) and do-it-yourself. Finally, in groups G and 12 we find event oriented customers with strong bidding activity in the tickets and travel category and in group H and 13, we find people bidding on sports equipment. 
     
Let us now focus on the time development of the user interests. The data for this analysis was collected during only a relatively short time span (25 days) and we base our results on an extremely sparse data set. Remember that every bidder in the network took part in only $3$ auctions on average. Is it really possible to predict meaningful patterns of consumer interest from such sparse data? One could further argue that the few most active bidders account for a large portion of the bids, thus holding the network together and ``defining'' the clusters of interest, because they also contribute a large number of links. In order to address this question, we revisited the data set in beginning of September 2005, more than nine months after our original study. From the 6 largest clusters of the $\gamma=1$ ordering, we uniformly and randomly sampled $10,000$ users each. Note that this removes possible bias towards very active users, they are now represented in the data according to their proportion in the population. Then we looked at the trading history of these users as far back as eBay permits - 90 days. For these $60,000$ users, we determined the product categories of the articles they had bought between June and September. Again, we calculated the odds ratios, this time of \textit{buying}, \ie winning an auction, from a particular category and with the new sample of users as basic population. The results are shown on the right hand side of Figure \ref{Clusterings} with a dashed background and the cluster id from which the users were sampled in parenthesis. The stability of the interest profiles is quite remarkable. The main interests have remained unchanged as compared to the initial study though in some cases the spectrum has become more diverse. For instance the content oriented bidders of cluster 1 now also show increased buying activity in the PC-games and tickets category. At the same time the main interest has shifted from movies to music. The largest number of product categories with increased odds of bidding in this category is found for cluster 9, the members of which are the most technology affine users anyway and which would be expected to satisfy a very broad range of consumer needs from online vendors. The members of cluster 7 (the collectors) and cluster 4 (the toy model builders) are much more conservative and almost do not change their profile at all. Without second hand data about the age structure of the bidders classified, we can only speculate that these clusters are formed by older customers who tend to stick to particular categories.  

\section*{Conclusion}
We have presented a detailed study of the user behavior on the online auction site \texttt{www.ebay.de} during the pre-Christmas season of 2004. Power-law distributed activity in terms of the number of articles sold, bought, and bid on was found. The attractiveness of articles, measured in terms of the number of bidders participating in an auction, shows an exponential distribution. Focussing on the bidding behavior, we constructed a network of bidders from their competition for single articles. Nodes in the network correspond to bidders and links to the fact that these bidders have expressed a common interest in at least one article. Studying the general statistical properties and comparing to appropriate random models, we find clear indications for a non trivial cluster structure. This cluster structure, its hierarchy and overlap was studied using a community detection algorithm. Our analysis did not need the definition of any kind of similarity measure between articles or product categories. Rather, we solely used the taxonomic information about articles provided by eBay to interpret our results. We can classify $85\%$ of the users into only a small number of well separated, large clusters, all of which have a distinct profile of only a few main interests as revealed by annotating the articles in the taxonomy of product categories. Some of the clusters show sub-clusters or overlap with other clusters. The interest profiles we identified are remarkably stable. Sampling randomly from the clusters and checking, what these users bought during a three month period in the summer 2005, we found that the profiles of articles bought were almost identical to those from the classification 6 months earlier.  

This is striking because virtually everything is offered on eBay and one would expect users to satisfy a much broader range of shopping interests. However, it appears that the major clusters mainly correspond to people's favorite spare time activities.  We believe the apparent stability of user's buying and bidding behavior reflects the permanence of their interests   which is also stabilized by their social environment and activities. The clear signature in the market data may stem from the fact that users tend to buy online only articles where they have some experience and expertise in. Users seem hesitant to bid on articles from categories in which they have not previously bid in. This may be due to the fact that inexperienced users cannot judge what is a fair price for an article in an auction and they have difficulty assessing to what extent the article offered really suits their needs. At the same time, user's interests are reinforced by online recommender systems \cite{Resnick1, Grouplens1}, which suggest similar articles to those already bought by the user. This temporal stability  corroborates the hypothesis that the presence of latent interest profiles in the society per se leads to the emergence of user groups with common interest. Transparent markets such as online auction sites in which users act independent and anonymously are perfect starting points for research into this collective behavior.  

\bibliographystyle{pnas}
\bibliography{../../../../BibTex_Citations}

\clearpage
\section*{Supporting Material}
\subsection*{User Activity}
\begin{figure}[ht]
\includegraphics[width=8 cm]{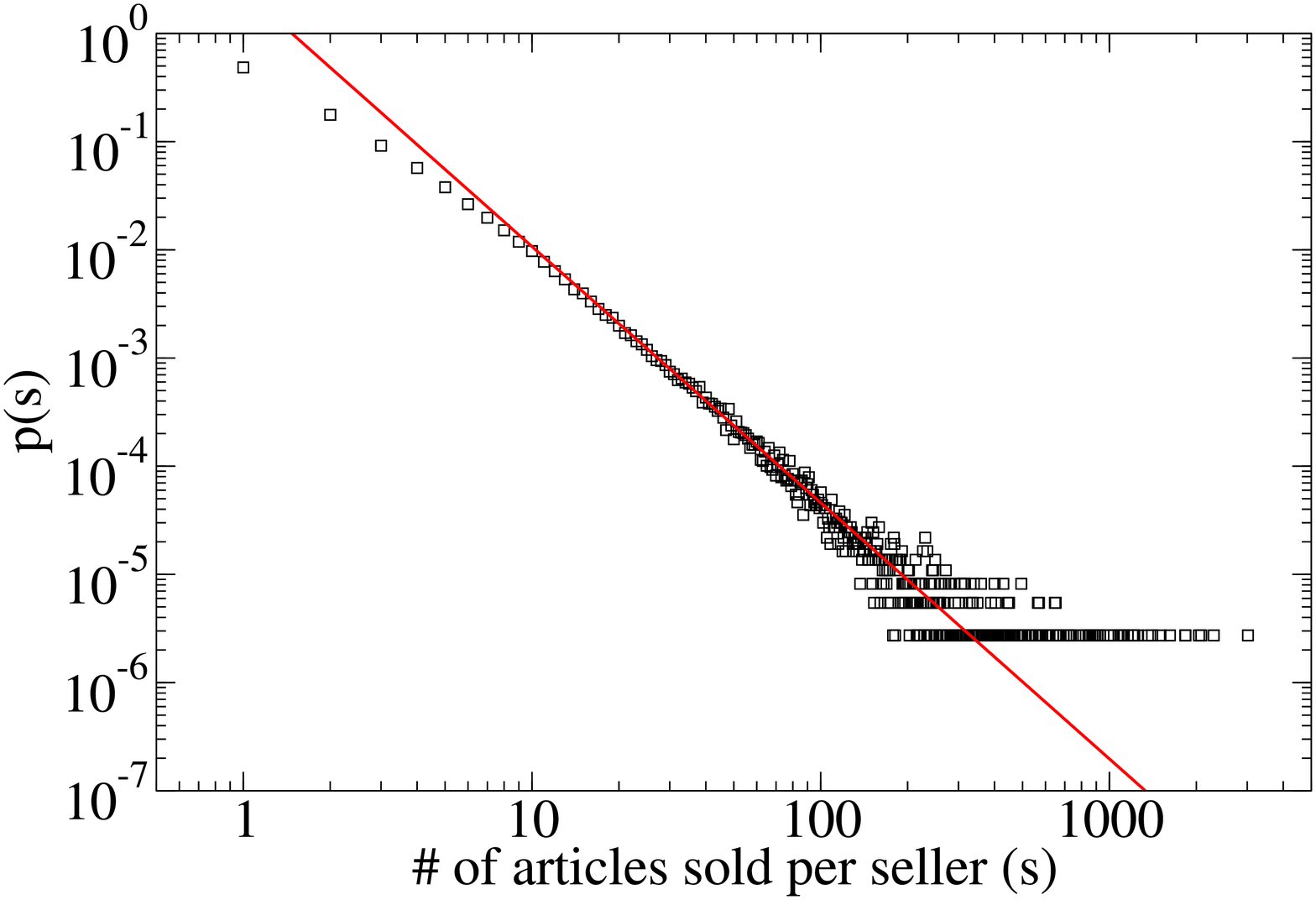}
\includegraphics[width=8 cm]{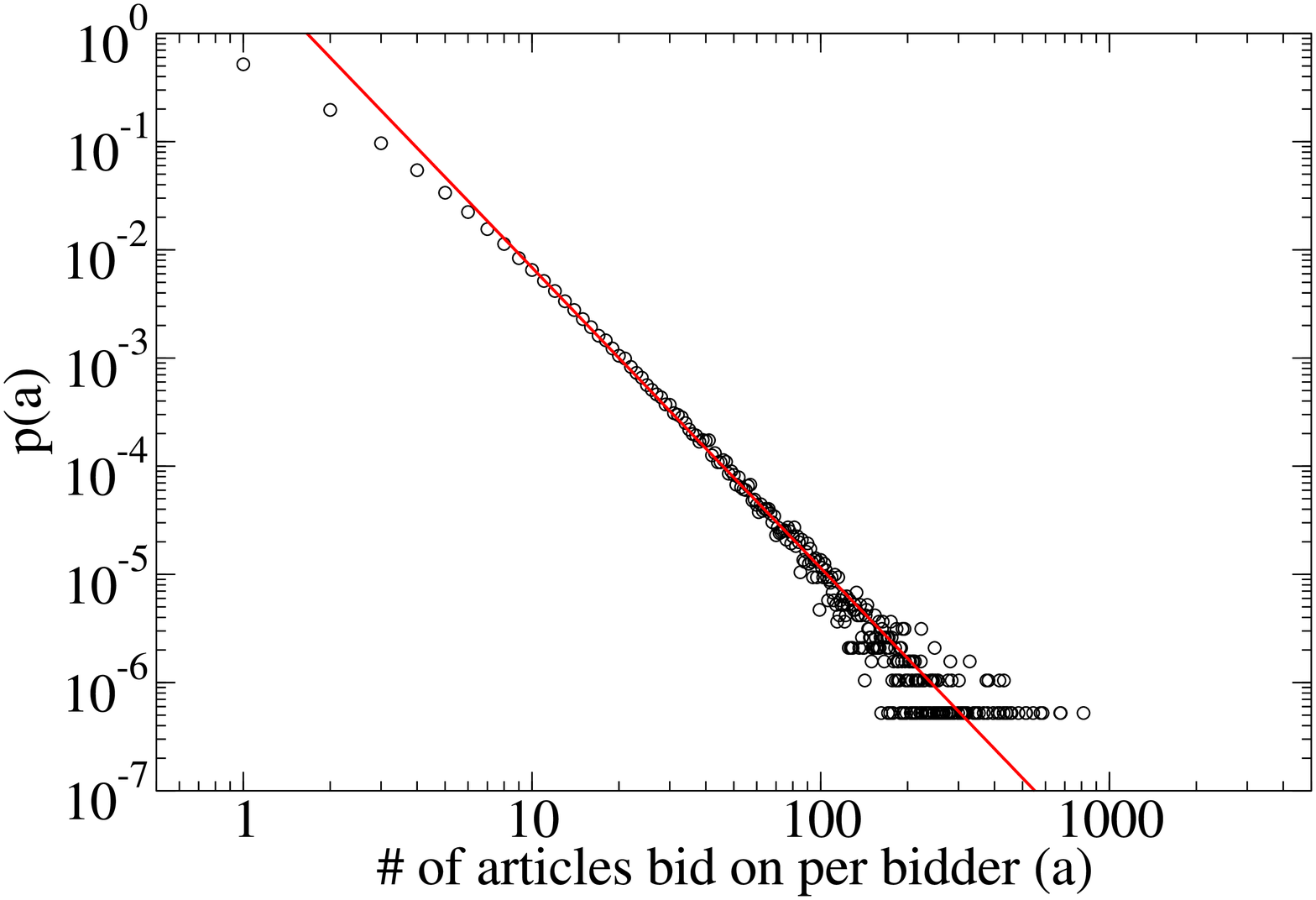}\\
\includegraphics[width=8 cm]{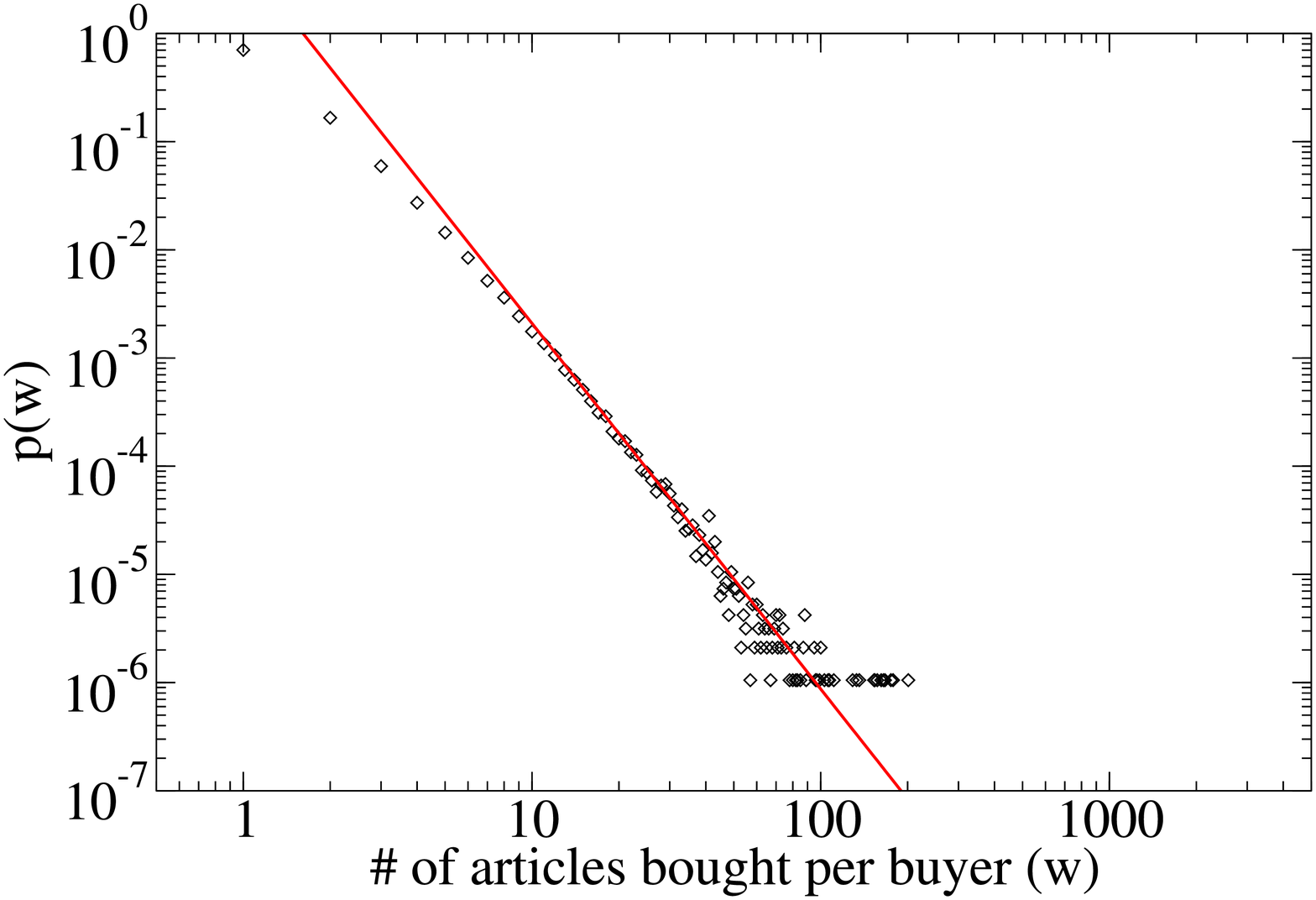}
\includegraphics[width=8 cm]{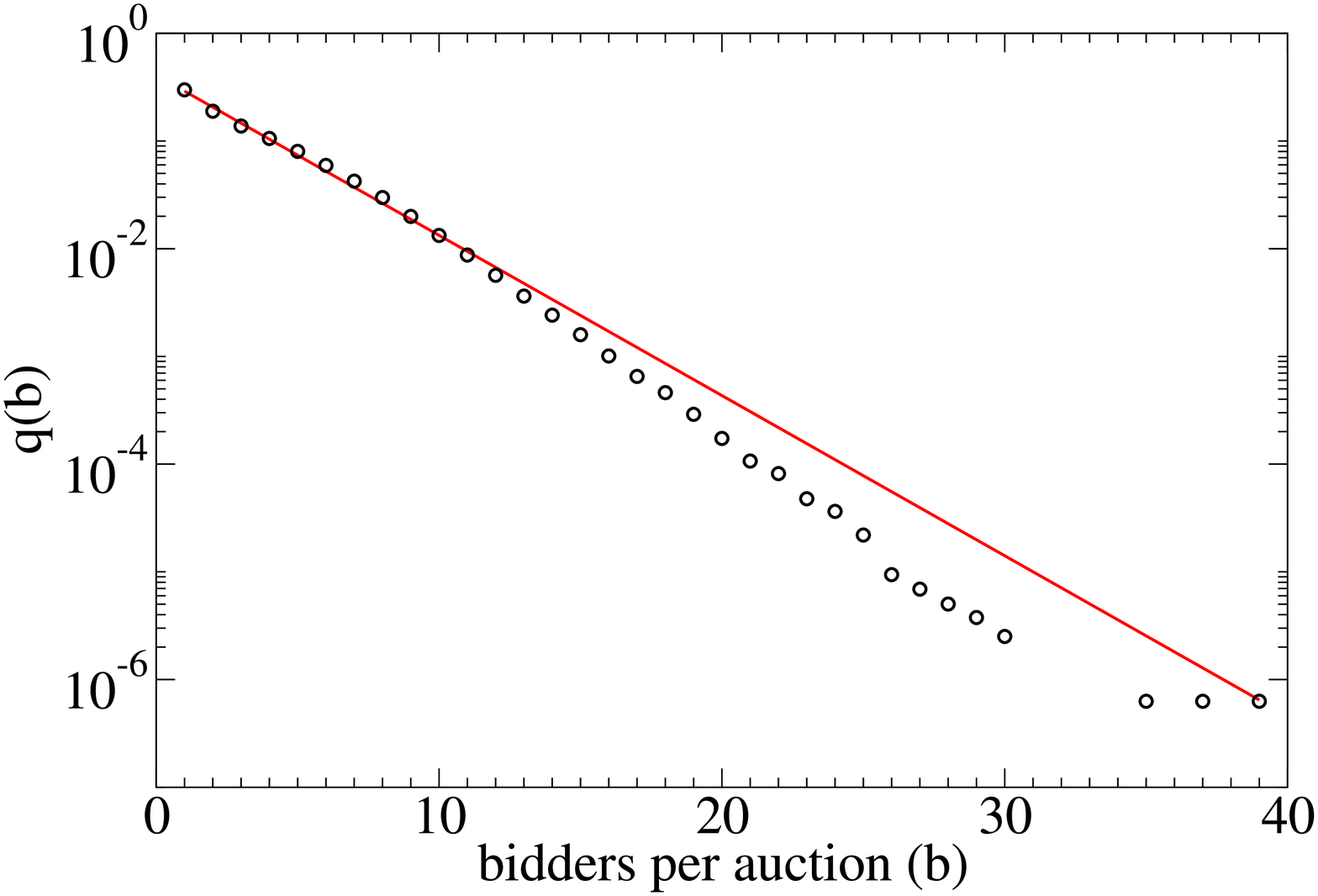}\\
\caption{User activity during the pre-Christmas season 2004. From top left to bottom right: probability mass function of the number of articles sold $p(s)$, different auctions participated in $p(a)$ and number of articles bought $p(w)$, and number of bidders participating in an auction $q(b)$. For the first three curves, the solid lines correspond to maximum likelihood fits of the tail of the distribution for $s,a,w\geq10$ with power-laws $p(x)\propto x^{-\kappa}$ and exponents  $\kappa_s=-2.37$, $\kappa_a=-2.78$ and $\kappa_w=-3.38$, respectively. For the distribution of the desirability  of an article $q(b)$, the dashed line represents a maximum likelihood exponential fit of the form $q(b)\propto \alpha^b$ with $\alpha=0.71$.}
\label{UserActivity}
\end{figure}

\noindent The distribution of the number of articles sold per seller falls off slowest, followed by the number of articles bid on and the number of articles bought. Here, we see the professionalization on the seller side of the market. There are ``power-sellers'' making a living from selling via eBay, but there are hardly any ``power-buyers'' professionally buying on eBay. This shows that eBay is more of a selling platform than an actual trading site, where selling and buying activities would be more balanced. If we assume that the tail of the distribution of the number of articles sold per seller is representative for the ``firm size'' of these users and compare these to the long term statistics of firm sizes in the US given by Axtell \cite{Axtell}, we can confirm the power law tail of the distribution, but not the exponent of $\kappa=2$. Instead, we find $\kappa=2.37$ and thus the observed distribution falls off faster. We can only speculate on the reasons for this and further study is needed here to compare new and old economy. In an earlier study, Yang \etal had reported an exponent of  $\kappa=3.5$ for the distribution of the number of auctions a bidder takes part in from a data set obtained in 2001 \cite{Kahng1} and we found $\kappa=2.78$ in our data. If this discrepancy is the result of a trend and not due to the differences in the observed countries and sizes of the data set, and this trend holds also for the distribution of the seller's activity, then one may be able to observe a convergence towards the exponent of $\kappa=2$ known from the old economy. 
\clearpage

\samepage{
\subsection*{Cluster Parameter}
\renewcommand{\baselinestretch}{1.25}
\begin{table}[h]
	\begin{center}
	\begin{tabular}{|r|r|r|r|r|r|}
	\hline
		Cluster	&	N	&	$\langle k_{in}\rangle$	&	$\langle k_{out}\rangle$	&	$p_{in}$	&	$p_{out}$	\\
		\hline
		A	&	200630	&	10.2	&	3.4	&	5.1E-05	&	5.0E-06	\\
		\hline
		1	&	84699	&	10.3	&	4.0	&	1.2E-04	&	5.0E-06	\\
		\hline
		2	&	29323	&	9.0	&	5.2	&	3.1E-04	&	6.0E-06	\\
		\hline
		3	&	76182	&	10.1	&	4.1	&	1.3E-04	&	5.0E-06	\\
		\hline
		\hline
		B	&	102188	&	18.6	&	3.9	&	1.8E-04	&	5.0E-06	\\
		\hline
		4	&	44830	&	24.6	&	4.2	&	5.5E-04	&	5.0E-06	\\
		\hline
		5	&	26325	&	14.2	&	5.2	&	5.4E-04	&	6.0E-06	\\
		\hline
		\hline
		C	&	19915	&	14.1	&	4.3	&	7.1E-04	&	5.0E-06	\\
		\hline
		6	&	20020	&	14.5	&	4.3	&	7.3E-04	&	5.0E-06	\\
		\hline
		\hline
		D	&	124702	&	16.5	&	3.8	&	1.3E-04	&	5.0E-06	\\
		\hline
		7	&	74913	&	17.2	&	4.1	&	2.3E-04	&	5.0E-06	\\
		\hline
		8	&	41359	&	16.8	&	5.9	&	4.1E-04	&	7.0E-06	\\
		\hline
		\hline
		E	&	183313	&	15.4	&	4.2	&	8.4E-05	&	6.0E-06	\\
		\hline
		9	&	73722	&	13.4	&	6.5	&	1.8E-04	&	8.0E-06	\\
		\hline
		10	&	47937	&	17.5	&	5.9	&	3.7E-04	&	7.0E-06	\\
		\hline
		\hline
		F	&	74657	&	10.5	&	4.9	&	1.4E-04	&	6.0E-06	\\
		\hline
		11	&	62115	&	11.1	&	5.0	&	1.8E-04	&	6.0E-06	\\
		\hline
		\hline
		G	&	31337	&	11.0	&	6.0	&	3.5E-04	&	7.0E-06	\\
		\hline
		12	&	18835	&	11.8	&	6.1	&	6.3E-04	&	7.0E-06	\\
		\hline
		\hline
		H	&	19620	&	10.0	&	4.4	&	5.1E-04	&	5.0E-06	\\
		\hline
		13	&	18286	&	9.9	&	4.4	&	5.4E-04	&	5.0E-06	\\
		\hline
	\end{tabular}
	\end{center}
\caption{Summary of basic parameters for the major communities found in the bidder network (annotated as in Figure $3$). $N$ denotes the number of bidders in the cluster, $\langle k_{in}\rangle$ and $\langle k_{out}\rangle$ the average numbers of neighbors within the cluster an in the rest of the network, respectively. By $p_{in}$ and $p_{out}$ we denote the internal and external link density, respectively. The average link density in the network is $\langle p\rangle=1.9\times10^{-5}$.}
\label{ClusterSummary}
\end{table}
}
\clearpage

\samepage{
\renewcommand{\baselinestretch}{1}
\begin{sidewaystable}
	\begin{footnotesize}
	\tabcolsep1mm
	\begin{tabular}{||r||r|r|r|r||r|r|r||r||r|r|r||r|r|r||r|r||r||r||}
	\hline\hline
Category & \textbf{A} &1 & 2 & 3 & \textbf{B} & 4 & 5 & \textbf{C} & \textbf{D} & 7 & 8 & \textbf{E} & 9 & 10 & \textbf{F} & 11 & \textbf{G} & \textbf{H} \\
          \hline
          \hline
Music     & \textbf{5.6}&  \textbf{8.6 (12.0)}&     &         &      &         &      &  1.5&     &        &      &     &        &       &    &          &      &     \\
\hline
Books     & 2.7&  3.4 (2.5)&     &     (1.1)&     &         &      &     &     &        &      &     &        &       &    &          &      &     \\
\hline
Movies    & \textbf{12.1}& \textbf{22.1 (9.2)}&     &         &      &         &      &     &     &        &      &     &        &       &    &          &      &     \\
\hline
Beauty & \textbf{5.6}&     (1.5) & \textbf{29.3} &     (1.3) &     &         &      &     &     &        &      &     &    (1.7) &      &    &          &      &     \\ 
\hline
Baby      & \textbf{13.7}&              &      &  \textbf{41.0 (20.0)} &     &         &      &     &     &        &      &     &        &       &    &          &      &     \\
\hline
Toys      & 1.3&              &      & \textbf{14.2} (4.9)&  \textbf{5.0}&     (1.6)&  1.1&     &     &        &      &     &        &       &    &          &      &     \\
\hline
Toy Models &    &              &      &         &  \textbf{70.2}& \textbf{21.9 (28.0)}& \textbf{10.2}&     &     &        &      &     &        &       &    &          &      &     \\
\hline
Audio     &    &     (1.3) &     &         &      &         &      & \textbf{16.2}&     &        &   1.2& 3.4& 2.1(3.9)&      & 1.0&   1.4 (2.3)&     &     \\
\hline
Instruments &    &     (1.5)&     &         &      &     (1.9) &     &\textbf{144.3}&     &        &      &     &    (2.2)&      &    &          &      &     \\
\hline 
Antiques  &    &              &      &         &   1.2&  1.8 (1.1)&     &     &  \textbf{6.6} & \textbf{6.3} (3.7)&     &     &        &       &    &          &      &     \\
\hline
Stamps    &    &              &      &         &      &         &      &     & \textbf{88.6} &\textbf{118.0 (18.0)}&     &     &        &       &    &          &      &     \\
\hline
Coins     &    &              &      &         &      &         &      &     & \textbf{66.7}& \textbf{88.3 (18.0)}&     &     &        &       &    &          &      &     \\
\hline
Collectibles &    &              &      &         &      &         &      &     &  1.6& 1.7 (2.6)&     &     &        &       &    &          &      &     \\
\hline
Jewelry & 1.1&              &   1.3&         &      &         &      &     &  4.5& \textbf{5.3} (3.1)&     &     &        &       &    &          &      &     \\
\hline
Photo     &    &              &      &         &      &     (1.2)&     &     &  4.4&        &  \textbf{47.9}&  2.8& 1.5 (3.2)&      &    &      (1.4) &     &     \\
\hline
Computer  &    &              &      &         &      &         &      &     &     &        &      & \textbf{19.6} & \textbf{14.7 (12.0)}&   3.1&    &      (1.4)&     &     \\
\hline
Cons. Electr.&    &              &      &         &      &         &      &     &     &        &      &  \textbf{6.7}&\textbf{16.1} (4.3)&      & 2.0&   2.5 \textbf{(5.9)}&     &     \\
\hline
Mobile    &    &              &      &         &      &         &      &     &     &        &      & \textbf{25.2} & 1.3 (4.7) & \textbf{ 33.0}&    &      (2.1) &     &     \\
\hline
Games     &    &     (2.2)&     &     (1.1)&     &         &      &     &     &        &      &  3.3& 2.3 (3.4)&   1.4&    &          &      &     \\
\hline
Software  &    &     (1.2)&     &         &      &         &      &     &     &        &   1.5&  \textbf{5.1}& 2.7 (4.0)&      &    &      (1.5)&     &     \\
\hline
Business  &    &              &      &         &      &         &      &     &     &        &      &     &    (1.1)&      &\textbf{21.5}&  \textbf{16.9 (15.0)}&     &     \\
\hline
DIY       &    &              &      &         &      &         &      &     &     &        &      &     &    (2.2)&      & \textbf{17.9}&  \textbf{19.1 (7.0)} &     &     \\
\hline
Motors    &    &              &      &         &      &         &   1.4&     &     &        &      &  1.5&    (1.6)&      & 8.6&   \textbf{8.2 (8.9)}&     &     \\
\hline
Travel    &    &     (1.2)&     &         &      &         &      &     &     &    (1.7)&     &     &    (1.4)&      &    &          &  \textbf{58.2}&     \\
\hline
Tickets   &    &     (2.7)&     &     (1.7)&     &         &      &     &     &        &      &     &    (1.7)&      &    &          &  \textbf{42.9}&     \\
\hline
Sports    &    &     (1.3)&     &     (1.3)&     &         &      &     &     &        &      &     &    (2.0)&      &    &      (1.9)&  1.4& \textbf{16.0}\\
\hline
Office    &    &     (1.1)&     &         &      &         &      &     &     &        &      &  2.3& 1.1(2.4)&      & 1.9&   1.6 (2.0)&     &     \\
\hline
Deli Food   &    &     (1.2)&     &         &      &         &      &     &     &        &      &     &    (1.5)&      &    &      (2.0) &  9.0&     \\
\hline
Household &    &              &      &         &      &         &      &     &     &        &      &  3.7& 3.3 (4.1) &      & 1.1&      (1.8)&  1.8&     \\
\hline
Fashion   & 1.9&              &   2.7&  1.7 (4.1)&     &         &      &     &     &        &      &     &    (1.2)&      &    &          &   1.1&  2.3\\
\hline
Furniture &    &     (1.0)&  1.1&     (1.6)&     &         &      &     &  1.3&        &      &     &    (1.6)&      &    &          &   4.2&     \\
\hline
Animals   &    &     (1.2)&  1.0&     (1.3)&     &         &      &     &     &        &      &  2.0&    (1.9)&      &    &      (1.4)&     &     \\
\hline
\hline	
	\end{tabular}
	\caption{Odds Ratios for \textit{bidding} in one of the 32 main categories during the pre-Christmas season 2004. Shown are only values above $1$ signifying an increased interest in articles from this category. Values larger or equal $5$ are set in bold font. Not how the clusters from the $\gamma=1$ clustering are more specfic than those from the $\gamma=0.5$ clustering, e.g. there are less categories with an OR larger $1$ and those that are deviate stronger from $1$.  Also note how the overlap between cluster A and B is mediated via the toy category and the overlap between cluster D and E via the photo category. For the six largest clusters of the $\gamma=1$ clustering, we show the OR of \textit{buying} from the corresponding category during the summer 2005 as explained in the text.}
	\label{SummaryTable}
	\end{footnotesize}
\end{sidewaystable}
}
\clearpage

\end{document}